\documentclass[11pt]{article}
\newcommand{\Comment}[1]{{}}
\usepackage[textwidth = 450 pt, textheight = 630 pt]{geometry}
\usepackage{amssymb,euscript,amsmath,amsfonts}
\usepackage{tikz,tikz-cd,url}
\usetikzlibrary{decorations.pathreplacing,calligraphy}
\usepackage{xcolor}
\usepackage{graphicx}
\usepackage{color}

\definecolor{MyDarkBlue}{rgb}{0.15,0.15,0.45}
\usepackage[linktocpage=true]{hyperref}
\hypersetup{final,
colorlinks=true,
citecolor=MyDarkBlue,
linkcolor=MyDarkBlue,
urlcolor=MyDarkBlue,
pdfauthor={  N. Lambert}
pdftitle={Title},
pdfsubject={hep-th}
}

\usepackage[numbers,sort&compress]{natbib}
\usepackage{hypernat}
\usepackage{accents}

\newcommand{\nn}{\nonumber}
\newcommand{\ie}{{\it i.e.\;}}

\newcommand{\be}{\begin{equation}}
\newcommand{\ee}{\end{equation}}
\newcommand{\bea}{\begin{eqnarray}}
\newcommand{\eea}{\end{eqnarray}}

\numberwithin{equation}{section}

\begin{document}

\renewcommand{\thefootnote}{\fnsymbol{footnote}}

   \vspace{1.8truecm}

 \centerline{\LARGE \bf {\sc Duality and Fluxes in the Sen   }}
 
  \centerline{\LARGE \bf {\sc    Formulation of Self-Dual Fields}}
 \vspace{1cm}

 \centerline{
   {\large {\bf  {\sc Neil~Lambert}\footnote{neil.lambert@kcl.ac.uk}}}}

\vspace{1cm}

\centerline{{\it Department of Mathematics}}
\centerline{{\it King's College London }} 
\centerline{{\it  London WC2R 2LS, UK}} 
\vspace{1.0truecm}

 
\thispagestyle{empty}

\centerline{\sc Abstract}
\vspace{0.4truecm}
\begin{center}
\begin{minipage}[c]{360pt}{

In Sen's formulation of  self-dual fields  one finds two closed forms: $H^{(g)}$ and $H^{(s)}$. Only the former couples to sources and the spacetime metric.  The latter  has the wrong sign kinetic term but  decouples and hence might be regarded as an unphysical artifact. In this letter we   illustrate how an electromagnetic duality associated to the potential for $H^{(s)}$ gives rise to a T-like duality in the partition function for $H^{(g)}$. We then compute  the partition function on a $4k+2$-dimensional torus highlighting its dependence on the choice of flux of  $H^{(s)}$.   Lastly we compute the two-point function of Wilson Surface operators. 
}
\end{minipage}
\end{center}
\renewcommand{\thefootnote}{\arabic{footnote}}
\setcounter{footnote}{0}
\newpage 

\section{Introduction}

After half a century of study self-dual fields, such as  the enigmatic  chiral Boson,  are still a subject of fascination.  Of course the case of a free chiral Boson in two dimensions has been well-understood for just as long. But self-dual fields and in particular interacting self-dual fields, pop-up in a variety of contexts and dimensions $D=4k+2$, $k=0,1,2,...$. There has been a recurrent and ongoing interest in Lagrangian models for the free Abelian versions of these theories \cite{Siegel:1983es,Floreanini:1987as,Pasti:1995tn,Perry:1996mk,Belov:2006jd, Mkrtchyan:2019opf,Townsend:2019koy} (see \cite{Evnin:2022kqn,Arvanitakis:2022bnr} for relations between some of these approaches). As well as a variety of computations of the  partition function \cite{Imbimbo:1987yt,Henneaux:1988gg,McClain:1990sx,Verlinde:1995mz,Dolan:1998qk,Henningson:1999dm, Chen:2013gca}.  On the other hand there is a general understanding   that although non-Abelian generalisations exist they do not admit   Lagrangian descriptions, at least in any standard way. 

 Here we wish to discuss some aspects of the Sen formulation of self-dual fields \cite{Sen:2015nph,Sen:2019qit}. See   \cite{Andriolo:2020ykk}   for  a discussion of this approach and \cite{Andriolo:2020ykk,Vanichchapongjaroen:2020wza, Andrianopoli:2022bzr,Chakrabarti:2022jcb,Barbagallo:2022kbt} for various applications and extensions. A major benefit of this formulation is that it is both Lorentz invariant and quadratic in the fields. Thus one can in principle make all computations in the quantum theory.  For example  \cite{Chakrabarti:2020dhv}  considered a $T{\bar T}$ deformation and, of particular relevance here,  \cite{Andriolo:2021gen} computed the exact partition function.

In the Sen formulation of self-dual fields one introduces a $2k$ form $B$ along with a $2k+1$ form $H$ that is self-dual with respect to a flat reference Minkowskian metric $\eta$: $H=\star_\eta H$.\footnote{More properly these are `pseudo-forms' as their transformation under diffeomorphisms is non-standard.} In the absence of sources the action can be written as
\begin{align}
S[H,B] = \frac{1}{(2\pi)^{4k+1}}\int \frac12 dB\wedge \star_\eta dB - 2H\wedge dB +H\wedge {\cal M}(H)	\ ,
\end{align}
where $\cal M$ is a linear map from $\eta$-self-dual forms to $\eta$-anti-self-dual forms.
The equations of motion are such that
\begin{align}
H^{(s)} &= \frac12 dB + 	 \frac12 \star_\eta dB+ H \nn\\
H^{(g)} &= H -{\cal M}(H) \ ,
\end{align}
are both closed. Clearly $H^{(s)}$ is self-dual with respect to the reference metric $\eta$. Furthermore ${\cal M}$ is constructed so that $H^{(g)}$ is self-dual with respect to the physical spacetime metric $g$. Thus one finds two self-dual fields but with respect to different metrics.

A key part of this approach is  that  $B$ has the wrong sign kinetic term and hence is pathological. Nevertheless $B$, through the combination $H^{(s)}$, decouples from any  sources and the physical metric $g$ and hence its pathologies do not affect the dynamics of the physically meaningful field $H^{(g)}$. In particular one finds that in a Hamiltonian formulation the theory splits into two, mutually non-interacting, systems described by $H^{(s)}$ and $H^{(g)}$ respectively. One is therefore free to simply ignore the pathological dynamics associated to $H^{(s)}$.

This all works very well but one issue is that the physical degrees of freedom are contained in $H^{(g)}$ while $B$ is unphysical. However many interesting questions about chiral form fields  involve Wilson surface operators which can not be captured by the field strength $H^{(g)}$. In particular we could be interested in objects of the form 
\begin{align}
W_\Sigma = e^{\frac{i}{(2\pi)^{2k}}\int_\Sigma B}\ ,
\end{align}
where $\Sigma$ is a closed $2k$-cycle in spacetime (for detailed discussions in six-dimensions see, for example, \cite{Henningson:1999xi,Gustavsson:2004gj,Drukker:2020dcz}).
The problem here is that $B$ is unphysical and a priori not related to the physical field $H^{(g)}$. So what is the meaning of $W_\Sigma$? How can we capture non-local observables associated to $H^{(g)}$? 

One answer to this question is to restrict to the sector  $H^{(s)}=0$ where we can think of $B$ as a potential for $H$ and hence $H^{(g)}$. 
We also note that in the reduction of the M5-brane on a Riemann surface discussed in \cite{Andriolo:2020ykk} it was   found that the correct Seiberg-Witten dynamics was only reproduced in an analogous restriction to vanishing unphysical flux.  Furthermore in string field theory $B$ originates from a string field $\tilde \Psi$ which gives rise RR sources \cite{Sen:2015uaa,FarooghMoosavian:2019yke} and therefore does have a physical role. This raises the more general question of whether or not the $H^{(s)}$ degrees of freedom have a more physical role to play than meets the eye. In particular it would appear that we shouldn't simply discard $B$ and hence $H^{(s)}$. Indeed, although this is possible classically, in a complete quantum treatment this can not be done so simply.

Our mandate here is  to understand the role of $H^{(s)}$ quantum mechanically within a path integral formulation. 
In particular we want to rewrite the partition function in the form
\begin{align}
Z	= \int dh^{(s)}_A \tilde Z[h^{(s)}_A]\ ,
\end{align}
where
\begin{align}
\tilde Z[h^{(s)}]	= \int [dH]_{\rm n.z.} [dB] e^{iS[B,H]}\ ,\end{align}
and $n.z.$ denotes the non-zero mode contributions to the partition function whereas $h^{(s)}_A$ denotes the zero modes. In fact we will see that the integral  reduces to a sum over discrete fluxes:
\begin{align}\label{Zhsum}
Z	= \sum_{h^{(s)}_A} \tilde Z[h^{(s)}_A]\ .
\end{align}
Once this is done we can then  restrict to  $h_A^{(s)}=0$  where we can think of $B$ as a local potential for $H$. But one is free to make other choices too.
Thus here we want to impose constraints on the fluxes $h^{(s)}_A$ and understand the resulting partition functions. 
This interpretation is reminiscent of a relative field theory \cite{Belov:2006jd,Freed:2012bs} in that we find a different partition function for $H^{(g)}$ depending on the choice of fluxes of $H^{(s)}$.

Therefore to explore this question we propose to look at     sectors of the theory labelled by  the periods of $H^{(s)}$ over closed $2k+1$ cycles $\Xi^A$. Note that since $H^{(s)}$ is self-dual it is sufficient to only include half the set of all closed $2k+1$ cycles. In particular we only consider $\Xi^A$ that are Poincare dual to anti-self-dual harmonic forms $\omega_A^-$:
\begin{align}
h^{(s)}_A = \frac{1}{(2\pi l)^{2k+1}}\int_{\Xi^A} H^{(s)} =  \frac{1}{2(2\pi l)^{4k+2}}\int H^{(s)}\wedge \omega^-_A	\ ,
\end{align}
where the extra factor of $2$ in the denominator in the second expression comes from our  conventions for $\omega_A^-$.

The calculation of the full partition function was presented in detail in \cite{Andriolo:2021gen}. Here we will revisit that calculation, separating out the zero-mode contribution and re-interpreting it. It is worth noting that that computation of the Euclidean partition function works in an interesting and novel way that is unique to the Sen formulation. Rather than Wick-rotating the time coordinate we analytically extend the physical spacetime metric $g$ to a Euclidean one. One then finds that all the geometrical information about the Euclidean torus is contained in $\widetilde {\cal M}$. As a result we can maintain the constraint $H=\star_\eta H$ and furthermore the wrong sign kinetic term for $B$ is harmless as it remains oscillatory in the path integral. Instead the damping needed to tame the path integral comes from the imaginary part of $\widetilde {\cal M}$. 
 
 The rest of this letter is structured as follows. In Section 2 we show how the electromagnetic duality of the $B$-field in the Sen action leads to duality on the fluxes of $H^{(s)}$ that is reminiscent of T-duality. Our discussion here is similar to  \cite{Verlinde:1995mz}. In section 3 we review  the computation of the partition function on a $4k+2$-dimensional torus ${\mathbb T}^{4k+2}$ \cite{Andriolo:2021gen} but this time separating out the contribution of the fluxes $h^{(s)}_A$ so that we can isolate the contribution of a single flux section. In section four we outline the computation of correlation functions of Wilson surface operators and give explicit results for the case of a chiral Boson. Lastly in section five we give our conclusions and comments. 
 
 \section{Duality}
 
 The Sen action admits a duality where we trade $B$ for a dual $2k$-form $\tilde B$. To this end we introduce $F = dB$ and write
  	\begin{align}
S[H,B] = \frac{1}{(2\pi)^{4k+1}}\int \frac12 F\wedge \star_\eta F - 2H\wedge F +H\wedge {\cal M}(H)	 + \tilde B\wedge dF\ .
\end{align}
 As usual integrating out $\tilde B$ leads to  $dF=0$ and hence $F = dB$, giving us back the Sen action. However we can also write it as
 	\begin{align}
S[H,B] = \frac{1}{(2\pi)^{4k+1}}\int \frac12 F\wedge \star_\eta F - 2H\wedge F +H\wedge {\cal M}(H)	 - d\tilde B\wedge F\ ,
\end{align}
and integrate out $F$ to find
 \begin{align}\label{Fdual}
\star_\eta F = -d\tilde B -2H  \ .
 \end{align} 
Substituting this back in to the Sen action leads to 
\begin{align}
S[H,B] = \frac{1}{(2\pi)^{4k+1}}\int \frac12 d\tilde B\wedge \star_\eta d\tilde B - 2H\wedge d\tilde B +H\wedge {\cal M}(H)	\ ,
\end{align}
so we have $S[H,B] = S[H,\tilde B]$. This will remain true in the presence of sources for $H$. However it is curious to note that 
\begin{align}
H^{(s)} & = H + \frac12 dB + \frac12 \star_\eta dB\nonumber\\
 & = H + \frac12 F + \frac12\star_\eta F\nonumber\\
 & = -\left(H + 	\frac12 d\tilde B + \frac12 \star_\eta d\tilde B\right)\nn\\
 &\equiv - \tilde H^{(s)}
\end{align}
In particular, from  (\ref{Fdual}) we see that the condition
\begin{align}
H^{(s)}=0\ 	
\end{align}
corresponds to $dB=d\tilde B$. 
 
Next we observed that we can identify two conserved $2k+1$-forms
\begin{align}
\Pi_N  &=  -dB - 2H\nonumber\\
\Pi_T &= \star_\eta dB	\ ,
\end{align}
which are co-closed. The first is the Noether current associated to the conserved momentum conjugate to $B$ whereas the second is topological.  We note that
\begin{align}
H^{(s)} = \tfrac12 (\Pi_T-\Pi_N)	\ .
\end{align}
The associated charges are
\begin{align}
Q_N(\Xi) &= -\frac{1}{(2\pi  )^{2k+1}}\int_\Xi 	\star_\eta dB + 2H\nonumber\\
Q_T(\Xi)  &= \frac{1}{(2\pi  )^{2k+1}}\int_\Xi 	dB \ .
\end{align}
where $\Xi$ is a closed $2k+1$ cycle.

To obtain a non-trivial topological charge we 
must allow for configurations where $B$ is not globally defined.  When the spacetime is a torus  we can think of this as including winding modes into $B$; that is introducing a linear dependence of $B$ on the torus coordinates $x^\mu$, in analogy with  a chiral Boson. To enable this we note that the so-called large gauge transformations
\begin{align}\label{LGsym}
	B\to B+ \omega\ ,
\end{align}
are symmetries provided $d\omega=0$. In particular on the torus we allow for solutions where
\begin{align}
B(x^\mu + 2\pi ln^\mu)	= B(x^\mu ) + \omega \ ,
\end{align}
where $n^\mu$ is a unit vector along some direction. We then see that
\begin{align}
 \frac{1}{(2\pi)^{2k+1}}\int_{S^1\times\Sigma} 	dB &= \frac{1}{(2\pi)^{2k+1}}\int_{ \Sigma} B(x^\mu + 2\pi ln^\mu)	- \frac{1}{(2\pi)^{2k+1}}\int_{ \Sigma} B(x^\mu )   \nn\\
 & = \frac{1}{(2\pi)^{2k+1}}\int_{ \Sigma} \omega \ ,\end{align}
 where the $S^1$ factor comes from transporting $\Sigma$ around a loop $x^\mu \to x^\mu + 2\pi ln^\mu$. 
Thus a flux quantization condition amounts to a restriction on which large gauge transformations we allow. We choose to impose
\begin{align}
Q_T(\Xi)
\in R^{2k+1}{\mathbb Z}\ ,
\end{align}
where $R$ is a positive constant.

Next we introduce Wilson Surface operators of the form
\begin{align}
W_{p,\Sigma} = e^{\frac{ip}{(2\pi )^{2k}}\int_\Sigma B}	\ .
\end{align}
 These are   invariant under   gauge transformations $B\to B+d\Lambda$  but for $W_{p,\Sigma}$ to be well defined under large gauge transformations we require that
\begin{align}
	 \frac{p}{(2\pi)^{2k}}\int_\Sigma \omega \in 2\pi {\mathbb Z}\ .
\end{align}
This in turn tells us that
\begin{align}
p \in \frac{1 }{R^{2k+1}}{\mathbb Z}	\ .
\end{align}
 Since we identify $p$ as the eigenvalue of the momentum operator we find
\begin{align}\label{pconstraint}
	Q_N(\Xi)  \in \frac{1}{R^{2k+1}}{\mathbb Z} \ .
\end{align}
In other words the momenta associated to the periods of $B$ are quantized. 
 From which we learn that the periods of $H^{(s)}$  satisfy 
 \begin{align}\label{HsFlux}
 \frac{1}{(2\pi )^{2k+1}}\int_\Xi 	H^{(s)}	\in \frac12\left( {R^{2k+1}}{\mathbb Z}  \oplus \frac{1}{R^{2k+1}}{\mathbb Z} \right)\ .
 \end{align}
 
 Finally we observe that the duality above corresponds to swapping $\Pi_N$ and $\Pi_T$ and hence $R\to 1/R$ and plays a role analogous to T-duality. The vanishing of the periods of $H^{(s)}$   therefore corresponds to the condition
\begin{align}
Q_N(\Xi)=Q_T(\Xi)	\ ,
\end{align}
which is invariant under the duality.

 \section{The Partition Function Revisited}

\subsection{Preliminaries}

Let us start by giving the complete form of the action we are interested in including sources \cite{Sen:2015nph,Sen:2019qit}:
\begin{align}
   S  [H,B;J]   = &\frac{1}{(2\pi)^{4k+1}}	\int  \left(\frac12 d B\wedge\star_\eta d B -2H\wedge d B +   (H+J)\wedge {  \widetilde{\cal M}}(H+J) + 2H\wedge J- \frac12 J\wedge \star_\eta J \right)\ ,
\end{align}
and the corresponding partition function is
\begin{align}
Z [J]  = \int [dH][dB] e^{i S_{{\cal A}}[H,B;J] }\ .
\end{align}
Here the integral is over a ${\mathbb T}^{4k+2}$ torus with periodic coordinates $x^\mu\in[-\pi l,\pi l]$.

To evaluate the partition function we need to have introduce  a basis of self-dual $2k+1$-forms $\omega^A_+$ and anti-self-dual  $2k+1$-forms $\omega_A^-$  where $A = 1,...,\left(\begin{array} {c}4 k+1\\ 2k \end{array}\right)$. For concreteness we choose the explicit basis   ($i_1<i_2<...<i_{2k}$):
\begin{align}
	\omega_+^A = (1+\star_\eta)dx^0\wedge dx^1\wedge\ldots dx^{2k}\nonumber\\
\omega_A^- = \pm (1-\star_\eta)dx^0\wedge dx^1\wedge\ldots dx^{2k}	\ ,
\end{align}
where the sign is chosen such that
\begin{align}
\omega_+^A\wedge \omega_{-B}	 = 2\delta^A_B d{\rm vol}\; ,
\end{align}
and in this basis we can write
\begin{align}
	{  \widetilde{\cal M}}(\omega_+^A) = {  \widetilde{\cal M}}^{AB}\omega^-_B\ .
\end{align}

As stated in the introduction the role of $\widetilde {\cal M}$ is to ensure that $H - \widetilde {\cal M}(H)$ is self-dual with respect to the physical spacetime metric $g$. To compute the partition function we want to consider a Euclidean metric $g$ while keeping the reference metric $\eta$ Minkowskian (so we can still impose the constraint $H = \star_\eta H$).  Thus we Wick rotate the metric and not the coordinates, as advocated in \cite{Visser:2017atf}.
It was shown in \cite{Andriolo:2021gen} that $\widetilde {\cal M}^{AB}$ becomes complex but crucially the imaginary part of $\widetilde{\cal M}^{AB}$ is negative definite which 
  leads to convergent Gaussian integrals. Thus even though there is a wrong-sign field in the Lagrangian the path integral can be well-defined and convergent if we instead analytically continue the physical metric and not the spacetime coordinates.

Furthermore we introduce a $\left(\begin{array} {c}4 k+2\\ 2k \end{array}\right)$-dimensional basis of $2k$-forms $\omega^a$. For concreteness we make the explicit choice
\begin{equation}
\omega^a  =  
	dx^{\mu_1}\wedge \ldots \wedge dx^{\mu_{2k}} \ ,
\end{equation}
where the index $a=(\mu_1,...,\mu_{2k})$ with $\mu_1<\mu_2<...<\mu_{2k}$.
It is helpful to expand
\begin{align} dx^\mu \wedge \omega^a= K^{\mu a}{}_B \omega^B_+ + L^{\mu a B}{} \omega_{-B} \;,
\end{align}
for some $K^{\mu a}{}_B $ and  $L^{\mu a B}{}$.
In particular 
\begin{align}
K^{\mu a}{}_A & = \frac{1}{2(2\pi l)^{4k+2}}\int  dx^\mu \wedge \omega^a\wedge \omega_{-A}	\nonumber\\
 L^{\mu a A} & = -\frac{1}{2(2\pi l)^{4k+2}}\int  dx^\mu \wedge \omega^a\wedge \omega_+^{A}\ .
 \end{align}
From which we see that the non-vanishing values of $K^{\mu a }{}_B,L^{\mu a B}$ are $ \tfrac12$ or $-\tfrac12$. However the non-vanishing values of $K^{\mu  a B}+ L^{\mu a}{}_B$ and $K^{\mu  a B}- L^{\mu a}{}_B$ are $1$ or $-1$.

To continue we expand the fields and sources  as
\begin{align}
H & = H_A\omega^A_+\nonumber\\
B &  = B_a\omega^a\nonumber\\
 J & = J^+_{A}\omega^A_+ + J^{A}_-\omega^-_{A}\ ,
\end{align}
where $H_A$,  $B_a$, $J_A^+$ and $J_-^A$ are functions of $x^\mu$.
The action can now be written as
 \begin{align}\label{action}
S_{\cal A} [H,B;J ] &= \frac{2}{(2\pi)^{4k+1}}\int  \Big(- \frac12 ( K^{\mu a}{}_BL^{\nu b B}{} +K^{\nu b}{}_BL^{\mu a B}{})\partial_\mu B_a\partial_\nu B_b -2L^{\mu a A}  H_A \partial_\mu B_a\nonumber\\
&\qquad\qquad\qquad\qquad +  (H_A+J^+_{A})(H_B+J^+_{B})\widetilde{\cal M}^{AB} +2H_A J^{A}_-  +J^+_{A}J^{A}_-  
\Big)\ .
\end{align}

We also find it helpful to introduce (note that this is a slightly different convention to \cite{Andriolo:2021gen}: ${\cal J}^A_{here} = \widetilde {\cal M}^{AB}{\cal J}^{there}_B $)
\begin{align}
{\cal J}^A = J^A_- +\widetilde {\cal M}^{AB}J^{+}_B \ .
\end{align}
Integrals are over a $4k+2$ torus ${\mathbb T}^{4k+2}$ and sums are over integers.  

\subsection{Re-Evaluation of the Partition Function}

Next we expand the functions $H_A$ and $B_a$ in terms of constant modes, winding modes and oscillating modes:
\begin{align}
B_a &= b_a + {w_{\mu a}}{} x^\mu +\sum'_{n_{a\mu}} b_{a,n_{a\mu}}e^{in_{a\mu} x^\mu /l}\qquad	n_{a\mu} \in {\mathbb Z}\ ,\nn\\
H_A & = h_{A} +\sum'_{n_{\mu}} h_{n_\mu A}e^{in_{\mu} x^\mu /l}\ ,
\end{align}
with  $(b_{a,n_{a\mu}})^*= b_{a,-n_{a\mu}}$ and $(h_{A,n_{a\mu}})^*= h_{A,-n_{a\mu}}$. Furthermore the winding modes take the form
\begin{align}
w_{\mu a} = \left(\frac{R}{ l}\right)^{2k+1}m_{\mu a}\qquad m_{\mu a}\in {\mathbb Z}	\ ,
\end{align}
for some constant $R$. Note that for $k>0$ not all $w_{\mu a}$ are independent, \ie\ different choices lead to the same flux of $dB$. A way to remove the redundancy is to only allow for non-vanishing $w_{\mu a}$ if $\mu<\mu_1$ where $\omega^a = dx^{\mu_1}\wedge dx^{\mu_2}\wedge...\wedge dx^{\mu _{2k}}$ with $\mu_1<\mu_2<...<\mu_{2k}$. The number of such $w_{\mu a}$ is then the same as the number of $2k+1$ forms. 

Since   the action is quadratic in the fields the oscillating  modes decouple from the zero-modes and winding modes and hence we can write:
\begin{align}
Z[J]  =e^{\frac{2i}{(2\pi)^{4k+1}} \int  \left( \widetilde{\cal M}-\widetilde{\cal M}^*\right)^{-1}_{AB}{\cal J}^A({\cal J}^B-\bar {\cal J}^B)}	Z_{\rm z.m.}[J]Z_{\rm osc.} \ .
\end{align} 

The oscillator modes of $B$ and $H$ are exactly as in \cite{Andriolo:2021gen}.
 Thus, as far as the oscillator modes are concerned, we can use the results of \cite{Andriolo:2021gen} which leads to 
\begin{align}
    Z_{\rm osc}&\sim \sqrt{\det\widetilde {\cal M}}\prod'_{n_\mu}\frac{1}{ \det (2K^{\mu a}{}_BL^{\nu b B}{}n_\mu n_\nu+2L^{\mu a A}L^{\nu b B}\widetilde{\cal M}^{-1}_{AB}n_\mu n_\nu)}
    \ .
\end{align}
The extra factor of $\sqrt{\det\widetilde {\cal M}}$ in $Z_{\rm osc.}$ compared to  \cite{Andriolo:2021gen} comes from the fact that full integral over $H$ gives simply 1 but here we have not integrated over the constant modes of $H$. 
Note that  $ Z_{\rm osc}$ also does not depend on    the sources $J$.
In analogy with the chiral Boson case we denote the regularised version of this expression by\footnote{We should also carefully deal with gauge redundancies, for example by fixing the gauge.}  \begin{align}
 Z_{\rm osc.}   = \frac{\sqrt{\det\widetilde{\cal M}}}{\eta_{4k+2}(\widetilde{\cal M}^{-1}_{AB})}	\ .
 \end{align}
 In particular for $k=0$ one finds that $\eta_2(\widetilde{\cal M}^{-1}) = \eta(-(1+\widetilde{\cal M}^{-1})/2)$  where $\eta$ is the   Dedekind $\eta$-function and $\widetilde{\cal M} = -(\tau+1)/(\tau-1)$ where $\tau$ is the usual modular parameter of a torus \cite{Andriolo:2021gen}.

Thus we are left with the zero-modes and winding modes. These contribute
\begin{align}\label{action}
S_{\rm z.m.}&= 4\pi l^{4k+2}\Big(  -\frac12( K^{\mu a}{}_BL^{\nu b B}{} +K^{\nu b}{}_BL^{\mu a B}{})w_{\mu a}w_{\nu b} -2L^{\mu a A}  h_A w_{\mu a}+ h_Ah_B\widetilde{\cal M}^{AB}\nonumber\\ 
&\qquad\qquad\qquad  + 2 h_AJ^+_{0B}\widetilde{\cal M}^{AB} +2h_A J^{-A}_0    
 \Big)\ .
 \end{align}
The main idea here is to make the substitution
\begin{align}
h^{(s)}_A = h_A +   K^{\mu a}{}_A w_{\mu a} 	\ ,
\end{align}
corresponding to the zero-mode of $H^{(s)} = H + \tfrac12( dB+\star_\eta dB)$.
Thus we have \begin{align}
S_{\rm z.m.}&= 4\pi l^{4k+2}\Big(  -\frac12( K^{\mu a}{}_BL^{\nu b B}{} +K^{\nu b}{}_BL^{\mu a B}{})w_{\mu a}w_{\nu b} -2L^{\mu a A} ( h^{(s)}_A-   K^{\nu a}{}_A w_{\nu a})w_{\mu a}\nonumber\\
&\qquad\qquad\qquad +  (h^{(s)}_A-  K^{\mu a}{}_A w_{\mu a} )(h^{(s)}_B-  K^{\nu b}{}_B w_{\nu b} )\widetilde{\cal M}^{AB}\nonumber\\ 
&\qquad\qquad\qquad + 2(h^{(s)}_A -  K^{\mu a}{}_A w_{\mu a})\widetilde{\cal M}^{AB}J^+_{0B}+2(h^{(s)}_A -  K^{\mu a}{}_A w_{\mu a})J^{-A}_0      \Big)\nonumber\\
&  =4\pi l^{4k+2}\Big(h^{(s)}_Ah^{(s)}_B\widetilde {\cal M}^{AB}  +K^{\nu b}{}_BL^{\mu a B}{}w_{\mu a}w_{\nu b}+K^{\mu a}{}_A K^{\nu b}{}_Bw_{\mu a}w_{\nu b}\widetilde {\cal M}^{AB}  \nonumber\\
&\qquad\qquad\qquad  -2h^{(s)}_A(L^{\mu a A} +   \widetilde {\cal M}^{AB} K^{\mu a}{}_A )w_{\mu a} +2(h^{(s)}_A -  K^{\mu a}{}_A w_{\mu a})(J^{-A}+\widetilde {\cal M}^{AB}J_{0B}^+)     \Big)\nonumber\\
& = 4\pi R^{4k+2}\Big((l/R)^{4k+2}h^{(s)}_Ah^{(s)}_B\widetilde {\cal M}^{AB} +\frac14 m^+_A m_-^A  +\frac14 m^+_A  m^+_B \widetilde {\cal M}^{AB}\nonumber\\
&\qquad\qquad \qquad  -(l/R)^{2k+1}h^{(s)}_A(m_-^A  +   \widetilde {\cal M}^{AB} m^+_B ) +(l/R)^{2k+1}(2(l/R)^{2k+1}h^{(s)+}_A -   m^+_A){\cal J}^{A}_0 \Big)\nonumber\\
& = \pi R^{4k+2}\Big(   ( m^+_A -s_A)( m^+_B -s_B)\widetilde {\cal M}^{AB} + ( m^+_A- 2s_A  )m^A_-  -4(l/R)^{2k+1}(    m^+_A-s_A){\cal J}_{0}^A \Big)\ .
\end{align}
Here we have introduced
\begin{align}\label{map}
m^+_A = 2K^{\mu a }{}_Am_{\mu a}\qquad m_-^A = 	2L^{\mu a A}m_{\mu a}\ ,
\end{align}
and
\begin{align}
{s}_A = 2\left(\frac{ l}{R}\right)^{2k+1}h^{(s)}_A	\ .
\end{align}

Since the non-zero values of $K^{\mu a }{}_A, L^{\mu a A}$ are $\pm \tfrac12$ $m^+_A$ and $m_-^A$  are integers. However since the non-zero values of $K^{\mu a }{}_A+L^{\mu a }{}_A$ are $\pm 1$    we see that  $m^+_A$ and $m_-^A$ are constrained so that 
\begin{align}
	m^+_A + m^A_- = 2 p^A\ ,
\end{align}
is even. Therefore we swap the sum over $n_{\mu a}$ for a sum over $m_A^+$ and $p^A$:
\begin{align}\label{action}
Z_{\rm z.m.} &=\int \prod_A dh^{(s)}_A \sum_{m^+_A,p^A}exp\Bigg( \pi i R^{4k+2}\Big(  ( m^+_A -s_A)( m^+_B -s_B)\widetilde {\cal M}^{AB}\nonumber\\
& \hskip4cm +  (m^+_A-2s_A  )(2p^A-m^+_A)   -4(l/R)^{2k+1}(    m^+_A -s_A) {\cal J}^A_{0}\Big)\Bigg)\nonumber\\
&=\int \prod_A dh^{(s)}_A \sum_{m^+_A,p^A}exp\Bigg( \pi i R^{4k+2}\Big(  ( m^+_A-s_A)( m^+_B -s_B)(\widetilde {\cal M}^{AB}-\delta^{AB})\nonumber\\
& \hskip4cm + 2(m^+_A-2s_A  )p^A +s_As_A  -4(l/R)^{2k+1}(    m^+_A -s_A) {\cal J}^A_{0}\Big)\Bigg)\ .
\end{align}
Note that the sum over $p^A$ enforces a $\delta$-function constraint
\begin{align}\label{constraint}
\sum_{p^A} e^{2\pi i R^{4k+2}(m^+_A-2s_A)p^A} =  \sum_{q_A} \delta(R^{4k+2}m^+_A-2R^{4k+2}s_A+q_A) \ .	 
\end{align}
Here we see that only sectors with quantised $h^{(s)}$ contribute to the path integral. In particular we only find contributions from
\begin{align}\label{sis}
s_A = \frac12 m_A^+ + \frac{1}{2R^{4k+2}}q_A	\ ,
\end{align}
or equivalently, noting that
\begin{align}
\frac{1}{(2\pi l)^{2k+1}}\int_{\Xi_B} \omega^+_A  =  2\delta_A^B 	\ ,
\end{align}
 from fluxes which satisfy the condition (for $R\ne 1$)
\begin{align}\label{Hflux}
	\frac{1}{ (2\pi  )^{2k+1}}\int_{\Xi_A} H^{(s)} \in\frac12 \left(R^{2k+1} \mathbb Z +\frac1{R^{2k+1}} \mathbb Z \right) \ ,
\end{align}
which agrees with the condition (\ref{HsFlux}) that we saw above. 

Thus we have put the partition function in the form of (\ref{Zhsum}):
\begin{align}
Z_{\rm z.m.} = \sum_{s_A}\tilde 	Z_{\rm z.m.} [s_A]\ ,
\end{align}
 in terms of  a sum of sectors of fixed $H^{(s)}$ flux. We could substitute $q_A$ for $s_A$ (\ref{sis})   in $Z_{z.m.}$ and find a double sum over the combination $m^+_A$ and $q_A$. This will result in the same expression for the partition function that was found in \cite{Andriolo:2021gen}. And in particular it was noted there that there is a symmetry $R\to 1/R$. 
However
the approach we want to take here is to restrict to certain sectors corresponding to   choices of $H^{(s)}$ flux. 

\subsection{Fixed flux}

To continue let us evaluate the contribution to the partition function from a fixed  value of $H^{(s)}$ flux where $s_A$ is given by
\begin{align}
s_A = \frac12 m_A + \frac{1}{2R^{4k+2}} n_A	\ ,
\end{align}
for a fixed  $m_A,n_A \in\mathbb Z$. An interesting case arises if $R^{4k+2}=r_1/r_2$ is rational where $r_1,r_2$ share no common divisors. In this case  there are infinitely many choices of $m^+_A$ and $q_A$   that solve the $\delta$-function constraint (\ref{constraint})
\begin{align}\label{mqsol}
m^+_A &= m_A + r_2 p_A\nn\\
q_A & = n_A - r_1p_A\ ,
\end{align}
for any $p_A\in \mathbb Z$. In this case we find
\begin{align}\label{tildeZis}
\tilde Z_{\rm z.m.}[s_A] &= \sum_{p_A}	exp\Bigg( \pi i \Big( r_1r_2(p_A +\tilde s_A/r_2)(p_B+\tilde s_B/r_2) (\widetilde {\cal M}^{AB}-\delta_{AB}) +r_1s_As_A/r_2 \nn\\ & \hskip7cm -4l^{2k+1}\sqrt{r_1r_2}(p_A+\tilde s_A/r_2){\cal J}^A_{0}  \Big)\Bigg)\nn\\
&= e^{\pi i  {r_1}{r_2}\tfrac{s_A}{r_2}\tfrac{s_A}{r_2}}\Theta \left[\begin{array}{c}
\tilde  s_A/r_2\\ 0	
\end{array}\right]
\left(-2l^{2k+1}\sqrt{r_1r_2}{\cal J}_{0}^A\mid r_1r_2(\widetilde{\cal M}^{AB}-\delta^{AB}) \right)\ ,
\end{align}
where 
\begin{align}
\tilde s_A  =  \frac12 m_A - \frac{1}{2R^{4k+2}} n_A	\ .
\end{align}
Here we have introduced the higher dimensional $\theta$-functions:
\begin{align}
  &\Theta \left[\begin{array}{c}
\alpha_A \\ \beta^A 	
\end{array}\right]
\left( z^A\mid {\cal T}^{AB} \right)   := \sum_{m_{A}}		e^{\pi i  (m_{A}+\alpha_A )(m_{B}+ \alpha_B ){\cal T}^{AB}+ 2\pi  i (m_{A}+\alpha_A)(z^A+\beta^A)}\ .
\end{align}
We see that for generic choices of flux we break  the duality $r_1\leftrightarrow r_2$. Although the sector with vanishing flux is invariant as expected:
\begin{align}
\tilde Z_{\rm z.m.}[0]  
&=  \Theta \left[\begin{array}{c}
0\\ 0	
\end{array}\right]
\left(-2l^{2k+1}\sqrt{r_1r_2}{\cal J}_{0}^A\mid r_1r_2(\widetilde{\cal M}^{AB}-\delta^{AB}) \right)\ .
\end{align}

 However we observe that
\begin{align}
 \frac{s_A}{r_2} = \frac{m_A}{2r_2} + \frac{n_A}{2r_1}\qquad 	\frac{\tilde s_A}{r_2} = \frac{m_A}{2r_2} - \frac{n_A}{2r_1}\  ,
\end{align}
and hence we can restore  the duality  by summing over sectors of fluxes which are symmetric under $m_A\leftrightarrow -n_A$, \ie if we include a flux $s_A = \tfrac12 m_A + \tfrac{r_2}{2r_1}n_A$ then we should also include $s'_A =  -\tfrac12 n_A -\tfrac{r_2}{2r_1}m_A$.
In particular if we sum over all fluxes sectors then the  winding mode contribution  to the full partition function is
\begin{align}
  Z_{\rm z.m.} &= \sum_{s_A}	\tilde Z[s_A]\nn\\
  &= \sum_{s_A}	e^{\pi i {r_1}{r_2}\tfrac{s_A}{r_2}\tfrac{s_A}{r_2}}\Theta \left[\begin{array}{c}
\tilde s_A/r_2\\ 0	
\end{array}\right]
\left(-2l^{2k+1}\sqrt{r_1r_2}{\cal J}_{0}^A\mid r_1r_2(\widetilde{\cal M}^{AB}-\delta^{AB}) \right)\ ,
\end{align}
where the sum is over all possible fluxes of the form (\ref{HsFlux}). This will again restore duality and give the result observed  in \cite{Andriolo:2021gen}.

\subsection{Self-dual and  Generic Radii}

Let us now  look at the radius $R=1$, corresponding to $r_1=r_2=1$. Here the fluxes take the simple values  
\begin{align}
	\frac{1}{(2\pi )^{2k+1}}\int_{\Xi_A} H^{(s)} \in \frac12 \mathbb Z \ ,
\end{align}
or $s_A\in \tfrac 12{\mathbb Z}$. This is somewhat degenerate from the discussion above as one can't distinguish between $m_A$ and $n_A$. As such the sum over $q_A$ in \ref{constraint} can be shifted to absorb $m^+_A$ and we obtain
\begin{align}
\tilde Z_{\rm z.m.}[s_A] 
& = e^{\pi i s_As_A}\Theta \left[\begin{array}{c}
 s_A\\ 0	
\end{array}\right]
\left(-2l^{2k+1} {\cal J}_{0}^A\mid  \widetilde{\cal M}^{AB}-\delta^{AB} \right)
\ .
\end{align}
Recall that $s_A$ is either   integer or half-integer. The integer parts can be absorbed in the sum over $m^+_A$ and so do not affect the $\Theta$-function but they can affect the phase. For example both even and odd $s_A$ lead to the same $\Theta$-function but the odd integer cases can come with  a minus sign.

Lastly we mention the case of a generic value of $R$, where $R^{4k+2}$ is not rational. Here there is a unique choice  $m^+_A=m_A$ and $q_A=n_A$  that solves the $\delta$-function constraint (\ref{constraint}) and we simply find
\begin{align}
\tilde Z_{\rm z.m.}[s_A]
= exp\Bigg( &\pi i R^{4k+2}\Big(  \tilde s_A\tilde s_B (\widetilde {\cal M}^{AB}-\delta^{AB})   -4(l/R)^{2k+1}\tilde s_A{\cal J}^A_{0}+s_As_A \Big)\Bigg)
\ .
\end{align}
 If we consider fluxes of the form (\ref{sis}) and sum over all $m_A$ with  $n_A$ fixed we'd again find a $\Theta$-function where  $n_A/R^{4k+2}$ would play the role of characteristics. Similarly if we summed over $n_A$ with $m_A$ fixed we'd find a $\Theta$-function with $m_A R^{4k+2}$ playing the role of characteristics. The duality $R\to 1/R$ would be broken.  Summing over both $m_A$ and $n_A$ will simply lead to $Z_{\rm z.m.} \sim 1$ as in \cite{Andriolo:2021gen} which of course trivially restores the duality.

\subsection{Non-integer Modings}

Reference \cite{Andriolo:2021gen}  introduced a constant $2k+1$ form ${\cal A}$ 
to allow for the flux condition
\begin{align}
\frac{1}{(2\pi )^{2k+1}}\int_{\Xi^A} (dB-{\cal A})  \in R^{2k+1}{\mathbb Z} 	\ ,
\end{align}
for some fixed $2k+1$ form  ${\cal A} = \alpha_{\mu a} dx^\mu \wedge \omega^a$.
In this case the winding modes are shifted away from integers to
\begin{align}
	w_{\mu a} = \left(\frac{R}{l}\right)^{2k+1}(m_{\mu a } + \alpha_{\mu a})\ .
\end{align}
This just corresponds to shifting $m^+_A\to m^+_A + \alpha^+_A$ and $m_-^A\to m_-^A + \alpha_-^A$ where $\alpha^+_A$ and $\alpha_-^A$ are obtained from $\alpha_{\mu a}$ using the map (\ref{map}). However one sees that the Wilson surface operators are no longer single valued on the torus and have a charge under the symmetry (\ref{LGsym}):
\begin{align}
W_{p,\Sigma}(B+\omega) = e^{2\pi i p \int_\Sigma \omega}W_{p,\Sigma}(B) \ .
\end{align}

In \cite{Andriolo:2021gen}, in order to arrive at a $\Theta$-function after summing over all fluxes, an extra topological term
\begin{align}\label{SA}
S_{\cal A} = -\frac{1}{(2\pi )^{2k+1}}	\int {\cal A}\wedge dB\ ,
\end{align}
 was included in the Lagrangian\footnote{Another boundary term was proposed in \cite{Chakrabarti:2022jcb} but that term vanishes here}.    The main effect of this term is to remove the $\alpha^+_A$ contribution that would otherwise appear in the  $\delta$-function (\ref{constraint}) so that we still   find  the fluxes are given by (\ref{sis}).
Thus the $H^{(s)}_A$ remain in the form (\ref{Hflux}). Repeating the calculation of $\tilde Z[s_A]$  now gives,
for $R^{4k+2}=r_1/r_2$,
\begin{align} 
\tilde Z_{\rm z.m.}[s_A] &=e^{\pi i \frac {r_1}{r_2}\left(s_As_A-\alpha^+_A(\alpha^+_A+\alpha_-^A)\right)} \Theta \left[\begin{array}{c}
  (\alpha^+_A+\tilde s_A)/r_2\\ r_1(\alpha^+_A+\alpha_-^A)	
\end{array}\right]
\left(-2l^{2k+1}\sqrt{r_1r_2}{\cal J}_{0}^A\mid r_1r_2(\widetilde{\cal M}^{AB}-\delta^{AB}) \right)\ .
\end{align}
 Note   that now the conjugate momentum is shifted:
 \begin{align}
 \Pi_N &=  (-dB-2H - \star_\eta{\cal A})	\nn\\
 \Pi_T &=  	(\star_\eta dB -\star_\eta {\cal A})\ ,
 \end{align}
 where we have also shifted the topological current to ensure again that $H^{(s)} = \tfrac12 (\Pi_T-\Pi_N)$.

 However from the point of view adopted here we do not need to add (\ref{SA}). 
Re-doing the calculation of $\tilde Z[s_A]$ without it   gives
 the  $\delta$-function constraint (\ref{sis}) 
\begin{align}
\sum_{p^A} e^{2\pi i R^{4k+2}(m^+_A+\alpha^+_A-2s_A)p^A} =  \sum_{q_A} \delta(R^{4k+2}m^+_A+R^{4k+2}\alpha^+_A-2R^{4k+2}s_A+q_A) \ . 
\end{align}
We see that a  non-zero $\alpha^+_A$ leads to a shift in the allowed fluxes for $H^{(s)}$:\begin{align} 
	\frac{1}{ (2\pi  )^{2k+1}}\int_{\Xi_A} (H^{(s)}-\tfrac12(1+\star_\eta){\cal A}) \in\frac12 \left(R^{2k+1} \mathbb Z +\frac1{R^{2k+1}} \mathbb Z \right) \ .
\end{align}
Let us consider the case $R^{4k+2}=r_1/r_2$ then for a given flux
\begin{align}
s_A = \frac12(m_A + \alpha^+_A)	 + \frac{r_2}{2r_1}n_A\ ,
\end{align}
we again find infinitely many solutions for $m^+_A, q _A$ of the form (\ref{mqsol}) and hence
\begin{align} 
\tilde Z_{\rm z.m.}[s_A] 
&= e^{\pi i \frac {r_1}{r_2}\left(s_As_A-s_A(\alpha^+_A+\alpha_-^A)-\alpha^+_A\alpha^+_A\right)}\nn\\
&\hskip 1cm \times \Theta \left[\begin{array}{c}
\tilde  s_A/r_2\\ r_1(\alpha^+_A+\alpha_-^A)/2	
\end{array}\right]
\left(-2l^{2k+1}\sqrt{r_1r_2}{\cal J}_{0}^A\mid r_1r_2(\widetilde{\cal M}^{AB}-\delta^{AB}) \right)\ .
\end{align}
 Similarly one can also find expressions for generic $R$.
 
\section{Wilson Surface Correlation Functions}

Let us now return to one of the motivations for this work, the computation of Wilson surface correlation functions. To this end we turn off the source $J$ and for simplicity we only consider integer moding. The extension to non-trivial moding should be straightforward. 

Our first task is to compute a one-point function
\begin{align}
	\langle W_{p,\Sigma}\rangle  = \int [dB][dH] e^{\frac{ip}{(2\pi)^{2k}}\int_\Sigma  B} e^{iS}\ .
\end{align} 
We start by evaluating
 \begin{align}
 	\frac{1}{(2\pi)^{2k}}\int_\Sigma  B & = l^{2k}b_a\Sigma^a +  l^{-1}R^{2k+1}x^\mu_\perp m_{\mu a}\Sigma^a+l^{2k}\sum b_{n_{\mu} a}\Sigma^ae^{i n_{\mu\Sigma}x^\mu_\perp}  \ ,
 \end{align}
where we have introduced  
\begin{align}
\Sigma^a = 	\frac{1}{(2\pi)^{2k}}\int_\Sigma \omega^a 	\ .
\end{align}
We also note that by our choice of convention for $n_{\mu a}$, $x^\mu n_{\mu a}$ is only non-zero for coordinates $x^\mu_\perp$ that are transverse to the cycle $\Sigma$. 

Thus we find three new terms in the action in addition  to those seen in the partition function computation above. The third term couples to the non-zero modes. However just as the non-zero mode contributions from the source cancel out in the path integral over $b_{\mu a}$ in the computation of the partition function \cite{Andriolo:2021gen},  these terms also do not contribute to the path integral when we integrate over $b_{\mu a }$. The second term   plays a similar role to that of the sources in the zero-mode calculation above and we will discuss them shortly.  

So let us first discuss the first term.  This comes from  a zero-mode $b_a$ that didn't explicitly enter into the action before. Nevertheless it was there but  the integral over it simply gave an overall infinite constant $\delta(0)$ which we discarded. However now we find
\begin{align}
\int db_a\Sigma^a e^{il^{2k}pb_a\Sigma^a}	= l^{-2k}\delta( p)\ .
\end{align}
This sets all the Wilson surface one-point functions to zero as expected by momentum conservation. 

To continue we can consider a two-point function:
\begin{align}
	\langle W_{p_1,\Sigma_1}W_{p_2,\Sigma_2}\rangle  = \int [dB][dH] e^{\frac{ip_1}{(2\pi)^{2k}}\int_{\Sigma_1}  B} e^{\frac{ip_2}{(2\pi)^{2k}}\int_{\Sigma_2}  B} e^{iS}\ ,
\end{align} 
where $\Sigma_1$ and $\Sigma_2$ are two $2k$-cycles related by translation along a transverse direction. For simplicity   put the second Wilson Surface at $x_\perp^\mu =0$ and the first at $x^\mu_\perp=y^\mu_\perp$. We therefore find  
\begin{align}
	\frac{p_1}{(2\pi)^{2k}}\int_{\Sigma_1}  B+\frac{p_2}{(2\pi)^{2k}}\int_{\Sigma_2} B & = l^{2k}(p_1+p_2)b_a\Sigma^a +  p_1l^{-1}R^{2k+1}y^\mu_\perp m_{\mu a }\Sigma^a+{\rm oscillators} \ .\end{align}
Again the integrals over the oscillator  modes  give the same result as without Wilson Surfaces. However  the integral over $b_a\Sigma^a$ now sets $p_1+p_2=0$ so in what follows we set $p_1=-p_2=p$.

Thus we are left evaluating the sum over winding modes:
\begin{align}
	\langle W_{p,\Sigma_1}W_{-p,\Sigma_2}\rangle&\sim   \sum_{m^+_A,m_-^A}{\rm exp}\Big( \pi i  R^{4k+2}  ( m^+_A -s_A)( m^+_B -s_B)\widetilde {\cal M}^{AB}\nn \\
	& \hskip3cm 
	+ \pi i  R^{4k+2} ( m^+_A- 2s_A  )m^A_- +  ipl^{-1} {R^{2k+1} }y^\mu_\perp m_{\mu a} \Sigma^a \Big) \ .
\end{align} 
Next we invert the relations in (\ref{map}) and write
\begin{align}
m_{\mu a} = {K'}^A_{\mu a} m^+_A	 + L'_{\mu a A}m^A_-\  ,
\end{align}
for some ${K'}^A_{\mu a}$ and $L'_{\mu a A}$.
Following as above we swap the sum over $m_{\mu a}$ for a sum over $m^+_A$ and $p_A$ where $m_A^- = 2p^A - m^+_A$
\begin{align}
	\langle W_{p,\Sigma_1}W_{-p,\Sigma_2}\rangle&\sim   \sum_{m^+_A,p^A}{\rm exp}\Big( \pi i  R^{4k+2}  ( m^+_A -s_A)( m^+_B -s_B)(\widetilde {\cal M}^{AB}-\delta^{AB})+\pi i  R^{4k+2} s_As_A\nn \\
	& \hskip3cm 
	+ 2\pi i   ( R^{4k+2}m^+_A- 2R^{4k+2}s_A + (\pi l)^{-1} p{R^{2k+1} }y^\mu_\perp L'_{\mu a A}\Sigma^a )p^A \nn\\
	&\hskip 3cm +  il^{-1}p {R^{2k+1} }y^\mu_\perp (K'^A_{\mu a}-L'_{\mu a A})\Sigma^a m^+_A \Big) \ .
\end{align}
The sum over $p^A$ in the second line gives a $\delta$-function
\begin{align}
\sum_{p^A}  &e^{2\pi i   ( R^{4k+2}m^+_A-   2R^{4k+2}s_A + (\pi l)^{-1} p{R^{2k+1} }y^\mu_\perp L'_{\mu a A}\Sigma^a )p^A}	\nn\\
&\hskip3cm= \sum_{q_A}\delta( R^{4k+2}m^+_A- 2R^{4k+2}s_A +q_A + (\pi l)^{-1}p {R^{2k+1} }y^\mu_\perp L'_{\mu a A}\Sigma^a )\ .
\end{align}
However for generic $y^\mu_\perp$ there is no solution   as the first three terms are discrete and the last continuous. In this way we find, for generic $y^\mu_\perp$,
\begin{align}
\sum_{p^A}  &e^{2\pi i   ( R^{4k+2}m^+_A- 2R^{4k+2}s_A + l^{-1}p {R^{2k+1} }y^\mu_\perp L'_{\mu a A}W^a )p^A}	\nn\\
& \hskip2cm = \sum_{q_A,q'_A}\delta_K( R^{4k+2}m^+_A- 2R^{4k+2}s_A +q_A-q'_A)\delta(q'_A +(\pi l)^{-1} p{R^{2k+1} }y^\mu_\perp L'_{\mu a A}\Sigma^a )
\nn\\
&\hskip2cm =
\sum_{q_A}\delta_K( R^{4k+2}m^+_A- 2R^{4k+2}s_A +q_A)\sum_{q'_A}\delta(q'_A +(\pi l)^{-1} p{R^{2k+1} }y^\mu_\perp L'_{\mu a A}\Sigma^a )
\ ,
\end{align}
where the first $\delta$-function is just the Kronecker $\delta_K$, \ie\ it takes the values $0,1$. Note that in the last line we shifted the sum over $q_A$ to absorb $q_A'$ since $q_A$ does not appear in the action. Thus we find a factorization and we identify
\begin{align}
 \sum_{q'_A}\delta(q'_A +(\pi l)^{-1}p {R^{2k+1} }y^\mu_\perp L'_{\mu a A}\Sigma^a )
  = \delta_{P_A}((\pi l)^{-1} p{R^{2k+1} }L'_{\mu a A}\Sigma^a y^\mu_\perp )\ ,
\end{align}
  as a periodic $\delta$-function with unit period.

The evaluation of the sum over $m^+_A$ works as it did above. In particular if we assume $R^{4k+2}=r_1/r_2$ then we find
\begin{align}
\langle W_{p,\Sigma_1}W_{-p,\Sigma_2}\rangle& =  \delta_{P_A}((\pi l)^{-1} p{R^{2k+1} }L'_{\mu a A}\Sigma^a y^\mu_\perp )\nn\\
&\times \Theta \left[\begin{array}{c}
\tilde  s_A/r_2\\ 0
\end{array}\right]
\left((2\pi l)^{-1} p\sqrt{r_1r_2}y^\mu_\perp (K'^A_{\mu a}-L'_{\mu a A} )\Sigma^a \mid r_1r_2(\widetilde{\cal M}^{AB}-\delta^{AB}) \right)\ .
\end{align}
 
This is   somewhat mired by notation,  so it is helpful to explicitly examine the case of a chiral Boson $k=0$. In this case the cycles $\Sigma_1,\Sigma_2$ are just  points and $y^\mu_\perp = (y^0,y^1)$.  Furthermore $a,A$ each  only take the value $1$ and a short computation shows that
\begin{align}
K^{01}_1=-K^{11}_1=\frac12 \qquad L^{011}=L^{111} = \frac12\ ,
\end{align}
and   
\begin{align}
K'^1_{01}=-K'^1_{11}=\frac12 \qquad L'_{011}=L'_{111} = \frac12\ .
\end{align}
Thus we simply find that 
\begin{align}
	(\pi l)^{-1}  K'_{\mu a A}\Sigma^a y^\mu_\perp  &= \frac{y^0-y^1}{2\pi l}\nn\\
	(\pi l)^{-1}  L'_{\mu a A}\Sigma^a y^\mu_\perp  &= \frac{y^0+y^1}{2\pi l}\ ,
\end{align}
are  the standard right and left moving coordinates and
\begin{align}
\langle W_{p,\Sigma_1}W_{-p,\Sigma_2}\rangle&=\delta_{P_1}(  p\sqrt{r_1/r_2}(y^0+y^1)/2\pi l)\nn\\
&\times \Theta \left[\begin{array}{c}
\tilde  s_1/r_2\\ 0
\end{array}\right]
\left( -p\sqrt{r_1r_2}y^1/2\pi l \mid r_1r_2(\widetilde{\cal M}^{11}-1) \right)\ .
\end{align}
Here $\widetilde{\cal M}^{11} -1= -2\tau/(\tau-1)$ where $\tau$ is the usual complex structure of the Euclidean torus.
Note that, due to the quantization condition $p R^{2k+1}\in {\mathbb Z}$, $p\sqrt{r_1/r_2}  $ and   $p\sqrt{r_1r_2} $ are both integers and hence we have invariance under the identification $y^\mu\to y^\mu + 2\pi l$ as we should. The periodic $\delta$-function then ensures that the correlation functions   depend on $y^0-y^1$.

\section{Conclusions}\label{conclusions}

In this letter we have explored the role that the fluxes of the unphysical $H^{(s)}$ field play in the quantum theory of the physical self-dual field $H^{(g)}$ in the Sen formulation of self-dual fields. We first showed that there is an electro-magnetic duality associated to the unphysical potential $B$ that leads to a T-like duality. We then re-derived the partition function from a path integral formulation but where the fluxes of the unphysical $H^{(s)}$ field were not integrated over. This gave us a family of partition functions for the physical field $H^{(g)}$ which depend on the choice of $H^{(s)}$ flux. In particular  the zero-mode part of the partition function obtained  from  a particular flux sector is given by (\ref{tildeZis}).  For a non-vanishing flux the duality is broken but can be restored by summing over pairs of fluxes.
We also computed the two-point function for Wilson surface operators and its dependence on the $H^{(s)}$ flux.

This picture is reminiscent  of a relative quantum field theory \cite{Belov:2006jd,Freed:2012bs} where there is not a unique partition function but rather the partition functions take values in the Hilbert space of an associated quantum field theory in one higher dimension. In our the associated  field theory is the free theory of the unphysical $H^{(s)}$ self-dual form.  In more well-known approaches the formulation of a self-dual field is obtained from a topological Chern-Simons type of action in $4k+3$ dimensions. It would be interesting to see if the Sen formulation has a similar origin (see \cite{Arvanitakis:2022bnr} for a recent discussion of such an approach for other actions). 

Although we have only considered the case of a toroidal spacetime we hope that much of what we have discussed can carry over to more general spacetimes. In particular the Wick rotation of the spacetime metric $g$, as first advocated in \cite{Andriolo:2021gen}, would appear to work well in any spacetime, leading to a  convergent path-integral despite having fields with a wrong-sign kinetic term. They simply remain oscillating. With this in mind it would be interesting to gain a better understanding of global features of the Sen formulation on more non-trivial spacetimes.

\section*{Acknowledgements}

I would like to thank E. Andriolo, D. Berman, C. Papageorgakis, A. Sen  and G. Watts for helpful discussions.

\bibliographystyle{utphys}
\bibliography{WSref.bib}

\end{document}